\documentclass{article}

\usepackage{arxiv}

\usepackage[utf8]{inputenc} 
\usepackage[T1]{fontenc}    
\usepackage{hyperref}       
\usepackage{url}            
\usepackage{booktabs}       
\usepackage{amsfonts}       
\usepackage{nicefrac}       
\usepackage{microtype}      
\usepackage{lipsum}
\usepackage{multirow}
\usepackage{mathbbol,bm,amsmath,amsthm,amsfonts,amssymb}
\usepackage{soul}
\usepackage{url}
\usepackage{caption}
\usepackage{lipsum}
\usepackage{color}
\usepackage{graphicx}

\DeclareMathOperator{\degr}{deg}
\DeclareMathOperator{\tr}{tr}
\newcommand{\R}{\mathbb{R}}
\newcommand{\I}{\mathbb{1}}
\newcommand{\A}{\mathbf{A}}
\newcommand{\D}{\mathbf{D}}
\newcommand{\X}{\mathbf{X}}
\newcommand{\Z}{\mathbf{Z}}
\newcommand{\U}{\mathbf{U}}
\newcommand{\V}{\mathbf{V}}
\newcommand{\W}{\mathbf{W}}
\newcommand{\g}{\mathbf{g}}

\newcommand{\Lbf}{\boldsymbol\Lambda}
\newcommand{\Dbf}{\boldsymbol\Delta}

\renewcommand{\L}{\mathbf{L}}
\renewcommand{\P}{\mathbf{P}}

\newcommand{\ecoli}{\textit{E.\ coli}}

\title{A Graph Feature Auto-Encoder for the Prediction of Unobserved Node Features on Biological Networks}

\author{
  Ramin Hasibi \\
  Department of Informatics\\
  University of Bergen\\
  Bergen, 5008 Norway \\
  \texttt{Ramin.Hasibi@uib.no} \\
   \And
  Tom Michoel \\
  Department of Informatics\\
  University of Bergen\\
  Bergen, 5008 Norway \\
  \texttt{tom.michoel@uib.no}  \\
}

\begin{document}
\maketitle

\begin{abstract}
\textbf{Motivation:} Molecular interaction networks summarize complex biological processes as graphs, whose structure is informative of biological function at multiple scales. Simultaneously,  omics technologies  measure the variation or activity of genes, proteins, or metabolites across individuals or experimental conditions. Integrating the complementary viewpoints of biological networks and omics data is an important task in bioinformatics, but existing methods treat networks as discrete structures, which are intrinsically difficult to integrate with continuous node features or activity measures. Graph neural networks map graph nodes into  a low-dimensional vector space representation, and can be trained to preserve both the local graph structure and the similarity between node features.
\\

\textbf{Results:}  We studied the representation of transcriptional, protein-protein and genetic interaction networks in \ecoli\ and mouse using graph neural networks. We found that such representations explain a large proportion of variation in gene expression data, and that using gene expression data as node features improves the reconstruction of the graph from the embedding. We further proposed a new end-to-end graph feature auto-encoder which is trained on the feature reconstruction task, and showed that it performs better at predicting unobserved node features than auto-encoders that are trained on the graph reconstruction task before learning to predict node features. When applied to the problem of imputing missing data in single-cell RNAseq data, our graph feature auto-encoder outperformed a state-of-the-art imputation method that does not use protein interaction information, showing the benefit of integrating biological networks and omics data using graph representation learning.
\\

\textbf{Availability:} The source code of this project is written in python and is available on \url{github.com/RaminHasibi/GraphFeatureAutoencoder}
\end{abstract}

\section{Introduction}

Biological networks of genetic, transcriptional, protein-protein, or metabolic interactions  summarize complex biological processes as graphs, whose structure or topology is informative of biological function at multiple scales. For instance, degree distributions reflect the relative importance of genes or proteins in a cell; 3-4 node network motifs have well-defined information-processing roles; and network clusters or communities contain genes or proteins involved in similar biological processes \cite{barabasi2004, zhu2007, alon2007b}. Simultaneously, genomics, transcriptomics, proteomics, and metabolomics technologies  measure the variation or activity of genes, proteins, or metabolites across individuals or experimental conditions \cite{ritchie2015methods,hasin2017multi}. There is a rich history of integrating the complementary viewpoints of biological networks and omics data. For instance, ``active subnetwork'' identification methods treat omics data as features of network nodes in order to identify well-connected subnetworks that are perturbed under different conditions \cite{nguyen2019comprehensive}. Network propagation or smoothing methods on the other hand use biological networks to extend partial information on some nodes (e.g., disease association labels, partially observed data) to other nodes (e.g., to discover new disease-associated genes or impute missing data) \cite{cowen2017network,ronen2018netsmooth}. However, existing methods treat biological networks as discrete structures, which are intrinsically difficult to integrate with continuous node features or activity measures. 

Recently, with the advent of deep learning, the idea of representation learning on graphs has been introduced. In this concept, nodes, subgraphs, or the entire graph are mapped into points in a low-dimensional vector space \cite{GraphRepSurvey1}. These frameworks are known as Graph Neural Networks (GNNs), and use deep auto-encoders to preserve the local structure of the graph around each node in the embedding, without having to specify in advance what ``local'' means. However, not much attention has been paid so far to the representation of the node features in these embeddings \cite{GraphRepSurvey3,chami2020machine}.

In this paper, we study whether graph representation learning of biological networks  using GNNs results in embeddings that are compatible with or informative for molecular profile data, concentrating for simplicity on gene expression data.  The three main contributions of this study are:
\begin{enumerate}
    \item  We introduce a method to systematically measure the relationship between the structure of a network and the node feature (gene expression) values. This is done using the Graph Auto-Encoder approach of \cite{GAE} and measuring \textit{(i)} the performance of reconstructing the network from the embedding, with and without node feautures, and \textit{(ii)} measuring the variance in feature values explained by the embedding matrix.
    \item We investigate how well GNNs preserve node features (expression values) in their graph representation, and introduce a new \emph{Graph Feature Auto-Encoder} layer

that is tailored to reconstructing the representation of the node features rather than the graph structure.

    \item We show that our new approach to graph representation learning has practical applications in tasks such as imputation of missing values in single cell RNA-seq data and similar scenarios.
\end{enumerate}

\section{Approach}
\label{sec:approach}

\subsection{Problem Formulation}
Assume that an undirected, unweighted graph $\mathcal{G}=(\mathcal{V},\mathcal{E})$ with $N=|\mathcal{V}|$ number of nodes (genes) has an adjacency matrix \textbf{A}, where $A_{ij}=1$ if there is an edge between nodes $i$ and $j$ and zero otherwise, and degree matrix \textbf{D}, a diagonal matrix with the degrees (number of neighbours) of each node on the diagonal. An $N\times Q$ matrix $\textbf{X}$ and $N\times P$ matrix $\textbf{Y}$, respectively called the feature matrix and the label matrix, denote the training and testing node feature values (molecular profiles), referred to as features and labels, of the $N$ nodes in $Q+P$ different experiments. We consider two approaches for mapping the graph into node feature values. In the first approach, the main purpose is to measure whether the graph structure and feature values encode similar information. In the second approach, we aim to reconstruct the feature matrix by leveraging the graph structure.

\subsection{Relationship between feature values and graph structure}
\label{sec:relat-betw-feat}

We learn a $N \times H$ node embedding matrix \textbf{Z} of the graph adjacency matrix \textbf{A} using a graph auto-encoder neural network:
\begin{equation}
\label{Eq:embedding_struct}
    \textbf{Z} = \textbf{GNN}(\textbf{X},\textbf{A}),
\end{equation}
as depicted in Fig. 1(A). We use the metrics of Average Precision and area under the ROC curve related to the reconstruction  of \textbf{A}, and the Variance Explained of $\textbf{Y}$ by \textbf{Z} quantify the relationship between the node features and graph structure.

\subsection{Prediction of unobserved node features}
\label{sec:pred-unobs-node}

\begin{figure*}
\centering
\begin{tabular}{@{}c@{}}
   \includegraphics[width=\columnwidth]{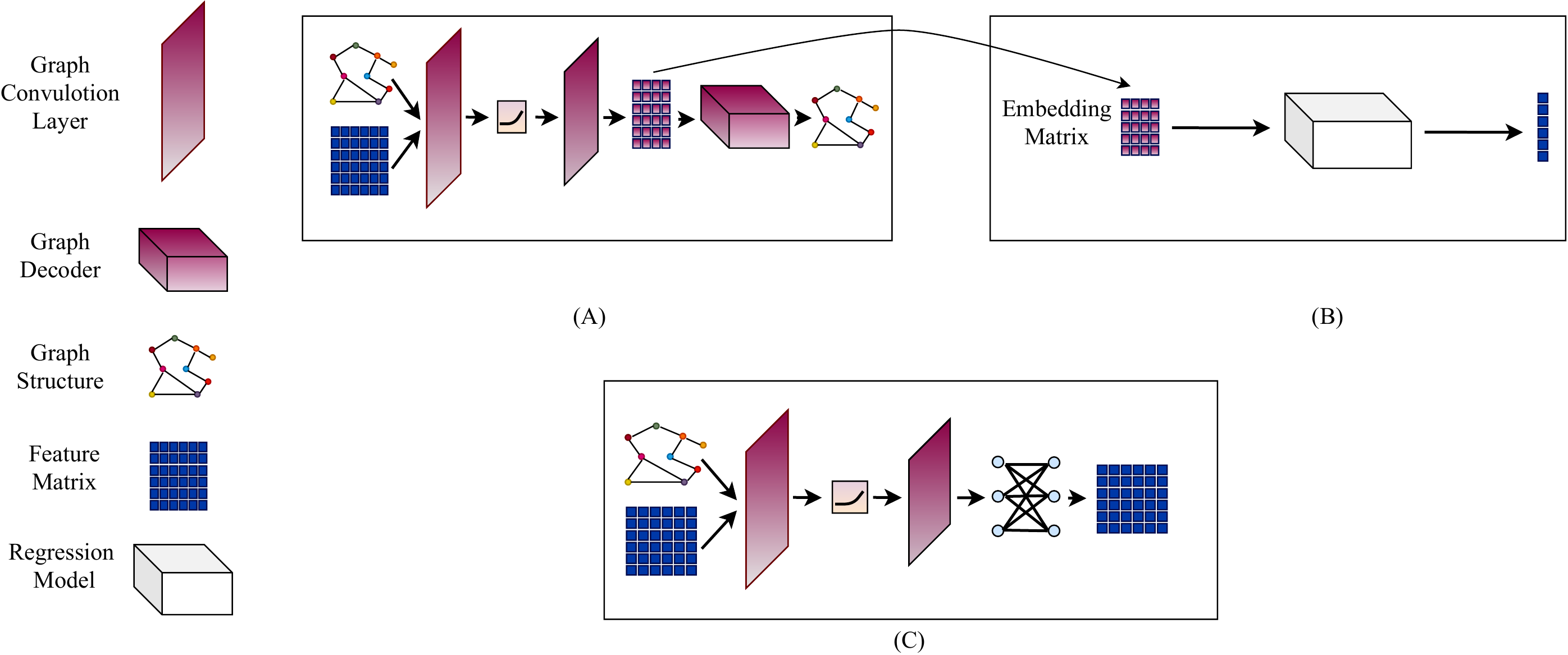} \\[\abovecaptionskip]
\end{tabular}
\caption{Different graph auto-encoder schemes. (A) depicts the graph auto encoder scheme of \cite{GAE} tailored to reconstruction of the adjacency of the graph. In (B) we take the embedding matrix of graph auto-encoder and train an indirect regression task for prediction of the expression values. (C) illustrates our proposed approach of graph feature auto-encoder for end-to-end learning of graph node features.}
\label{fig:auto-encoders}
 \end{figure*}

\subsection{Neural network on graphs}

One of the first attempts at learning neural networks over graph structures was the convolution operation on graphs.
For an input signal $x \in \mathbb{R}^N$,  the spectral convolution is defined as
\begin{equation}
    \g_\theta * x = \U \g_\theta \U^T x,
\end{equation}
in which $\U$ is the matrix of eigenvectors of the symmetric Laplacian $\L = \D-\A =  \U \Lbf \U^T$. $\U^Tx$ is called the Fourier transform of signal $x$ and $\g_\theta$ is a matrix function of $\Lbf$, the diagonal matrix of eigenvalues of $\L$.

Due to the high cost of calculating the  eigenvalues in the case of large matrices, \cite{chebyshevNets} proposed to use a Chebyshev series expansion truncated after the $K^{th}$ term to approximate the graph convolution operation with a $K^{th}$-order polynomial: 

\begin{equation}
\label{Eq:eq_cheb}
    \g_\theta * x \approx \U \sum_{k=0}^K \theta_k^{'} T_k (\tilde{\Lbf}) \U^T x=\sum_{k=0}^K\theta_k^{'} T_k(\tilde{\Lbf})x,
\end{equation}
in which $T_k(.)$ and $\theta_k^{'}$ are the $k^{th}$-order Chebyshev polynomials and expansion coefficients, respectively, $\tilde{\Lbf}=\frac{2}{\lambda_{max}}\Lbf- \I_N$ with $\lambda_{max}$ the largest eigenvalue of $\Lbf$, and $\tilde{\L} = \U \tilde{\Lbf} \U^T = \frac{2}{\lambda_{max} }\L- \I_N$. 

In Graph Convolution Networks (GCN) \cite{GCN}, further approximations were done by setting $K=1$, $\lambda_{max}\approx 2$, and $\theta = \theta_0^{'} = -\theta^{'}_1$. As a result, (\ref{Eq:eq_cheb}) was transformed into
\begin{equation}
\label{Eq:eq_GCN_filt}
    \g_\theta * x \approx (\I_n+\D^{-\frac{1}{2}}\A\D^{-\frac{1}{2}})x.
\end{equation}
Repeated application of $\g_\theta$ resulting in high powers of $\D^{-\frac{1}{2}}\A\D^{-\frac{1}{2}}$ can cause numerical instabilities.  \cite{GCN} suggested to set the diagonal elements of \textbf{A} to 1 (add self-loops) and to recompute \textbf{D} according to the updated adjacency matrix. Therefore, they used the symmetrically normalized adjacency matrix $\tilde{\textbf{A}}$ in their convolution layer, with
 \begin{equation}
    \tilde{\A} = \D^{-1/2}\A\D^{-1/2}.
  \end{equation}

Thus, the forward operation in a Graph Convulotion Network (GCN) for $Q$ input signals arranged in a $N\times Q$ matrix $\X$ is computed as
\begin{equation}
\label{Eq:GCN_forward}
    \Z = \sigma(\tilde{\A}\:\mathrm{ReLU}(\tilde{\A}\X\W_0)\W_1)
\end{equation}
with weight matrices $\mathrm{\textbf{W}}_i$ containing the trainable weights for each input feature, and $\sigma$ a non-linear task specific function such as softmax for a node classification problem \cite{GraphRepSurvey4}.

Additional studies on GNNs have show that a GNN can be viewed as a message-passing approach based on graph structure, where every node's message for its neighbours is the aggregation of the neighbourhood information, where the aggregation is done through a trainable neural network \cite{MessagePassing}. 
This framework is also known as a Message Passing Neural Network (MPNN). The forward pass in such a network consists of a message passing phase and a readout phate. In the message passing phase, the hidden representation of each node is updated through an update function, which aggregates the previous node representation and the messages of its neighbours according to :
\begin{equation}
    h_{i}^k=\gamma^k(h_{i}^{k-1},\sum_{j\in N(i)}M(h^{k-1}_i,h^{k-1}_j,e_{ij})),
\end{equation}
in which $h_i^k$ is the hidden state of node $i$ in layer $k$, with $h^0_i$ being the node $i$'s features and $e_{ij}$ is the edge attribute between nodes $i$ and $j$. Additionally, $\gamma$ and $M$ are both differentiable functions called the update and message functions, respectively. Furthermore, $N(i)$ denotes the set of neighbouring nodes of node $i$.

In the readout phase, the feature vector of the graph is computed using some learnable, differentiable readout function $R$ according to
\begin{equation}
    \textbf{Y} = \textbf{R}(\{h_i|i\in\mathcal{V}\}).
\end{equation}

\section{Methods}
\label{sec:methods}

\subsection{Graph representation learning for structural embedding}
\label{sec:graph-repr-learn}

To learn the representation of the structure of a graoh, we used the Non-probabilistic Graph Auto-Encoder model of \cite{GAE}. The embedding matrix $\Z$ is calculated by 
\begin{equation}
    \label{Eq:eq_z}
   \mathrm{ \Z = GCN(\textbf{X},\textbf{A})},
\end{equation}
where GCN(\textbf{X},\textbf{A}) is a two layer Graph Convolutional Network on the input \textbf{A} and \textbf{X} (the features of the graph). GCN(\textbf{X},\textbf{A}) is obtained by setting $\sigma$ in (\ref{Eq:GCN_forward}) to the identity function,
\begin{equation}
    \mathrm{GCN}(\X,\A) = \tilde{\A}\mathrm{ReLU(\tilde{\A}\X \W_0)\W_1}.
\label{Eq:eq_GCN}
\end{equation}
The weights are trained by measuring how well the embedding reconstructs the graph adjacency matrix, where the reconstructed adjacancy matrix $\hat{\textbf{A}}$ is defined as
\begin{equation}
    \hat{\textbf{A}} = \mathrm{Sigmoid}(\textbf{Z}\textbf{Z}^T).
\end{equation}
The cross-entropy error over all the edges in the matrix is used as a loss function,
\begin{equation}
    \label{Eq:loss}
    \mathcal{L} = -\sum_{n=1}^{N} \A_{n}\:\mathrm{ln}\:\hat{\A}_{n},
  \end{equation}
  in which $\A_{n}$ is the adjacency row of the $n$th node in $\A$ and $\tilde{\A}$.
The training of the neural network is done by gradient descent and stochasticity added by dropout rate.

\subsection{Message passing neural network for end-to-end prediction of features}
\label{sec:mess-pass-neur}

We used three popular message passing networks for finding the hidden representation of the nodes, as well as introducing our own, for the  task of predicting node feature values based on their network neighbourhood information. These three methods are inductive GCN, GraphSAGE  \cite{GraphSAGE}, and the GNN operator from \cite{GraphConv} (from here on out refered to as GraphConv). 
According to \cite{MessagePassing}, the formula for inductive GCN is
\begin{equation}
    h_i^k = \sum_{j\in N(i)\cup i}\frac{1}{\sqrt{deg(i)}*\sqrt{deg(j)}}.(\mathrm{W}h_j^{k-1}).
\end{equation}
The formula for GraphSAGE is
\begin{equation}
    h_i^k = \mathrm{W_1}(h^{k-1}_i) + \mathrm{W_2}\mathrm{Mean}_{j \in N(i)\cup i} (h^{k-1}_j).
\end{equation}
The GraphConv operator is calculated through
\begin{equation}
    h_i^{k} = \mathrm{W}_1h_i^{k-1}+\sum_{j \in N(i)}\mathrm{W}_2.h_j^{k-1}.
\end{equation}
Our version of MPNN  predicts the features of a node based on its own features and those of its neighbours. Therefore, we first obtain a representation of every node's features by running them through a linear layer. Then, we aggregate the neighbouring information by mean pooling of messages.  For the update function we concatenate the node features with its aggregated message representation, and run them through a shared weight network which determines how important each of these values are in predicting the features of the node. Hence, the formulae for our FeatGraphConv operator are as follows
\begin{align}
    g_i^{k} &= \mathrm{W_1} * h_i^{k-1}\\
    h_i^{k} &= \mathrm{W_2}(g_i^{k}||\mathrm{Mean}_{j \in N(i)\cup i} (g^{k}_j)),
\end{align}
in which $(.||.)$ is the concatanation function.

\subsection{Prediction of features from features}
\label{sec:pred-feat-from}

For comparison of the results obtained through GNNs, we also consider prediction of \textbf{Y} directly from \textbf{X} through simple machine learning algorithms. These algorithms include:
\begin{itemize}
    \item \textbf{Multi Layered Perceptron (MLP)}: A simple form of neural networks which maps the input features into output features through multiple layers of neurons (computing units).
    \item \textbf{Linear Regression:} A linear model for mapping the input to output.
    \item \textbf{Random Forest:} A set of decision tree models that determine the output value through the aggregation of output of decision trees that each are trained on a subset of \textbf{X}
    \item \textbf{Markov Affinity-based Graph Imputation of Cells (MAGIC)}: Uses signal-processing principles similar to those used to clarify blurry and grainy images to recover missing values from already existing ones in a matrix. \cite{Magic}
\end{itemize}

\subsection{Experimental setup}
\label{sec:experimental-setup}

\begin{table}[!t]
\centering
\caption{Hyper-parameters of the graph neural network}
\begin{tabular}{@{}lcc@{}}\toprule 
Hyper-parameter & Node Embedding & MPNN
\\\midrule
Epochs & 500 & 20000\\
initial learning rate & 0.001 & 0.001\\
first hidden layer size & 64 & 64\\
second hidden layer size & 32 & 32\\\hline
\end{tabular}
\label{Tab:01}
\end{table}

The hyper-parameters of all the experiments were determined after some initial experiments and were kept the same for all the models, to measure the predictive performance of different approaches under the same set of initial circumstances. These hyper-parameters are listed in Table \ref{Tab:01}. 

\subsubsection{Train-test splitting of features and nodes and edges}

We divided different properties of the graph into train and test sets depending on the experiment. When experimenting on the relationship between different graphs and the node features (cf.\ Section~\ref{sec:relat-betw-feat}), for the task of calculating the average percision and AUC of adjacency reconstruction we follow the work of \cite{GAE}, and split the $\mathcal{E}$ of each graph into 3 folds, 2 of which are for training and the remaining is used for testing. For calculating the Variance Explained and Average Mean Square Error (MSE) of prediction of each dimension of \textbf{Y}, we split the nodes in 3 folds, 2 of which are used for training and the remaining for testing. As a result, a regression model is trained on the embedding of the training nodes for each of the features (experiments) and is used to predict or measure the Variance Explained on the feature for the test nodes (Fig 1.a).

For assessing the feature auto-encoder scheme (cf.\ Section~\ref{sec:pred-unobs-node}), we randomly divide all the features values into 3 separate folds. The aim is to reconstruct the training values and measure the error on the test values (Fig 1.b). After every run the folds are exchanged in training and testing, such that the error measured is the mean of all the test errors in 3 different runs. This procedure is also known as K-fold cross validation.

\subsubsection{Structure embedding}
\label{sec:graph-embedd-meth}

We used the PytorchGeometry implementation of the graph auto-encoder provided by \cite{PytorchGeo}. For our approach, the normal auto-encoder provided in the package was used, and the variational auto-encoder was omitted. 
Four different sets of input graphs and features to the model were tested:
\begin{enumerate}
    \item \textbf{Random Graph}: In this approach $\Z$ is calculated by
    \begin{equation}
        \Z = \mathrm{GCN}(\I_N,\A_{rand}),
    \end{equation}
    in which, $\I_N$ and $\A_{rand}$ represent  an identity matrix of size $N \times N$ and the adjacency matrix of a random graph, respectively. For generating random graphs, we used the random graph generator of the Python3 package NetworkX, using the Erd\H{o}s–R\'enyi model \cite{RandomGraph} (Appendix)    
    \item \textbf{Expression + Random Graph}: In this approach, the identity matrix is replaced with the actual expression values of genes as input features:
    \begin{equation}
        \Z = \mathrm{GCN}(\mathbf{X},\A_{rand}).
    \end{equation}
    \item \textbf{Real Graph}: In this approach, the embedding matrix $\Z$ is calculated by
    \begin{equation}
        \Z = \mathrm{GCN}(\I_N,\A),
    \end{equation}
    where \textbf{A} is the adjacency matrix of the input graph.
    \item \textbf{Expression + Real Graph}: The embedding \textbf{Z} in this case is calculated through (\ref{Eq:eq_z}),
      \begin{equation*}
        \Z = \mathrm{GCN}(\mathbf{X},\A).
    \end{equation*}
    \item \textbf{Expression}: The network in this model, is inferred from the correlation between the expression values of different genes. In this approach, the correlation directly outputs the probability of the edge between two nodes.
\end{enumerate}
By choosing the identity matrix as input features in approach 1 and 3,  each one of the nodes has a distinct set of features which do not have any indication to the functionality of the node. This way the model will only pay attention to the graph structure when producing the embedding matrix.

\subsubsection{Node feature representation learning}
Since \textbf{Y} has $P$ number of dimensions, we can either train a different model for prediction of each of the dimensions, or predict \textbf{Y} in an auto-encoder manner. In the latter, the values of \textbf{Y} are considered to be missing values in the feature matrix and the model is trained on reconstruction of the \textbf{X} values. In the first approach, the evaluation of the methods is done on the prediction of each of the \textbf{Y} dimensions of the test genes. In the second approach, the \textbf{Y} values are scattered in the feature matrix, similar to the case of imputation problems. The difference between these two methods is illustrated in Fig. 2.
\begin{figure}
\centering
      \includegraphics[width=0.8\columnwidth]{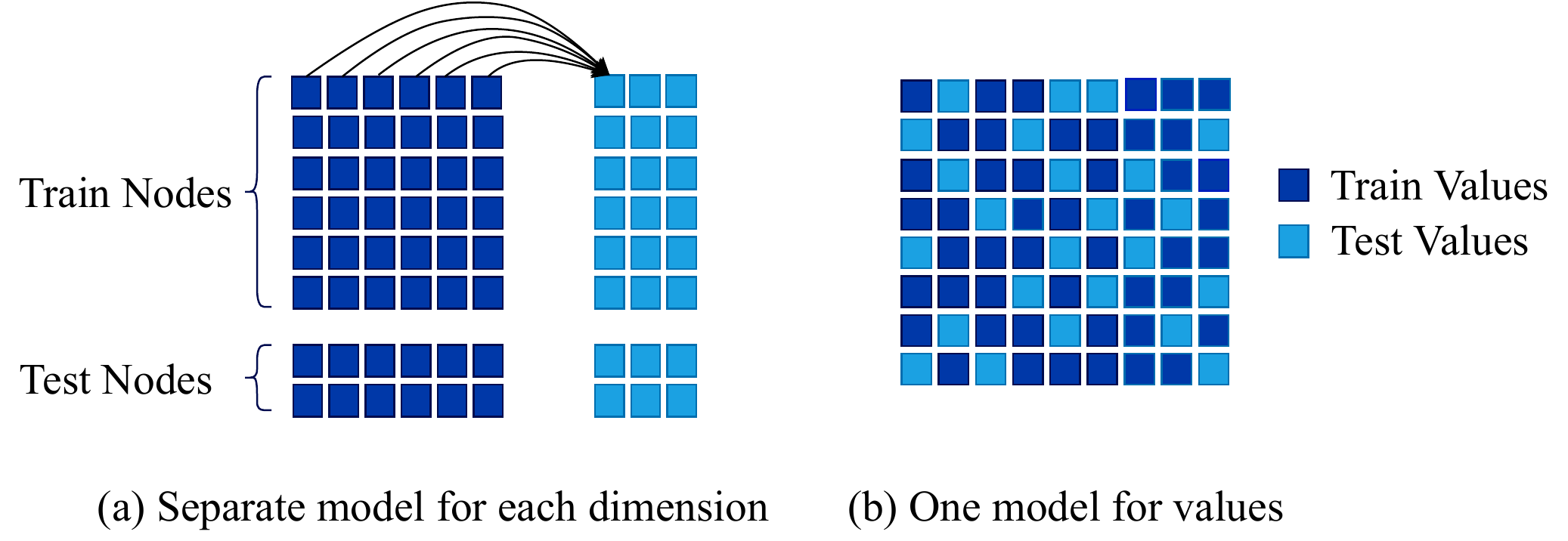} \\[\abovecaptionskip]
      \caption{The two different approaches of prediction of expression values. In (a), each dimension of \textbf{Y} is predicted using a separate model trained on \textbf{X}. In (b), a single model is used to predict all the values of the feature matrix.}
      \label{fig:train-test}
\end{figure}

For the MPNN modules of inductive GCN, GraphSage, and GraphConv we used the already available implementations from the package PytorchGeo. Furthermore, we used MessagePassing class of this package to implement our proposed FeatGraphConv. The hyper-parameters of all the MPNNs are available in Table \ref{Tab:01}. The architectures of all the models consist of two message passing layers followed by a fully connected layer as the readout function, with Relu as the activation function of the message passing layers. As before, both the cases of with node features and without node features are used in the training of the MPNN. In the case of nodes without features,  $\I_N$ is used, turning (\ref{Eq:eq_cheb}) into 
\begin{equation}
     \mathrm{Y} = \mathrm{GNN}(\I_N,\A).
\end{equation}

\subsubsection{Datasets}
\label{sec:datasets}

We evaluated the performance of our method on data for the organisms \textit{Escherichia Coli} (\textit{E.\ coli}) and {\textit{Mus Musculus}}. 
\begin{itemize}
    \item\textbf{Network datasets:} For \textit{E.\ coli}, we used transcriptional, protein-protein, and genetic interaction networks. The transcription network was obtained from RegulonDB \cite{regDB:2019}.  All the positive and negative regulatory effects from the TF-gene interactions dataset file were included regardless of their degree of evidence (strong or weak) to construct the adjacency matrix. A PPI and genetic interaction network were obtained  from BioGRID \cite{BioGRID_data,BioGRID_Paper}. We extracted the interactions from the file ''BIOGRID-ORGANISM-Escherichia\_coli\_K12\_W3110-3.5.180'', and considered the ''physical'' and ''genetic'' values of 'Experimental System Type' column for constructing the PPI and genetic networks, respectively. For \textit{Mus Musculus}, we used a protein-protein interaction network extracted in the same way from the file  ''BIOGRID-ORGANISM-Mus\_musculus-3.5.182''.
    \item \textbf{Expression level dataset:}  For \textit{E.\ coli}, we used the Many Microbes Microarray Database (M3DB) \cite{M3db_data,M3db_paper}. All the experiments from the file ''avg\_E\_coli\_v4\_Build\_6\_exps466probes4297'' were used to construct the feature matrix. For {\textit{Mus Musculus}}, the single cell RNA-Seq data from \cite{SingleCellData} were obtained from the Gene Expression Omnibus. 
\end{itemize}
For each of the networks, the common genes between the network and the expression data were extracted, and an adjacency matrix and a matrix of features were constructed from the network and expression level datasets, respectively. A detailed description of each of the networks is available in table \ref{Tab:02}.

\begin{table}[!t]
\centering
\caption{Summary description of benchmark datasets}

\begin{tabular}{@{}cclccclccc@{}}
\toprule 
\multirow{2}{*}{Input} & \multicolumn{2}{c}{\multirow{2}{*}{Expressions}} & \multicolumn{2}{c}{TF\_net} & \multicolumn{3}{c}{PPI}               & \multicolumn{2}{c}{Genetic} \\
                       & \multicolumn{2}{c}{}              & Nodes        & Edges        & \multicolumn{2}{c}{Nodes} & Edges     & Nodes        & Edges        \\ \hline
Ecoli                  & \multicolumn{2}{c}{466}                          & 1559         & 3184         & \multicolumn{2}{c}{1929}  & 11592     & 3688         & 147475       \\ \hline
Mus Musculus           & \multicolumn{2}{c}{1468}                         & -            & -            & \multicolumn{2}{c}{9951}  & 75587     & -            & -           
\end{tabular}
\label{Tab:02}
\end{table}

\subsection{Computational resources and Source Code}
All the experiments were done on Tesla V100 with Python 3. The source code of the experiments is availabe at \url{github.com/RaminHasibi/GraphFeatureAutoencoder}

\enlargethispage{6pt}

\section{Results}

\subsection{Measuring the relationship between biological networks and gene expression values}
\label{sec:meas-relat-betw}

To explore whether the structure of biological networks relates to the expression values of  genes, and whether such a relationship can be expressed quantitatively using graph neural networks, we learned low-dimensional embeddings of  transcriptional regulatory (TF\_net), protein-protein interaction (PPI) and genetic interaction networks in \ecoli\ and a protein-protein interaction network in mouse, with and without expression data as node features, and trained to optimize reconstruction of the original graph from the node embedding (see Methods Section \ref{sec:methods} for details). Table \ref{tab:AUC} shows the area under the ROC curve (AUC) and Average Precision results for the graph reconstruction task for various embeddings. Embeddings learned from the structure of the real graph alone (``Graph'' row) performed considerably better than embeddings learned from random graphs (``Random Graph'' row), as expected. The same was true for a standard Pearson correlation coexpression network inferred from the expression data alone (``Expression'' row), showing that graph embeddings and gene expression data independently predict graph structure. Interestingly, when gene expression data were used as node features during the graph representation learning (see Section~\ref{sec:graph-repr-learn} and \ref{sec:graph-embedd-meth}), graph reconstruction performance further increased (``Expression $+$ Graph'' row), but this was not the case when expression data was combined with random graphs (``Expression $+$ Random Graph'' row). In other words, graph embeddings where the distance between nodes respects both their graph topological distance and their expression similarity results in better graph reconstruction than embeddings that are based on topological information alone. This shows that expression profiles are informative of graph structure in a way that is consistent with, but different from, the traditional view where networks are inferred directly from expression data using expression similarity measures.

Next we computed the variance of the expression data explained by the different embeddings (see Appendix \ref{sec:vari-expl-feat}). Despite being trained  on the graph reconstruction task, graph embeddings learned with and without expression data as node features explained a high percentage of variation in the expression data, but not when random graphs were used (Table~\ref{tab:VE}).

In summary, graph representation learning results in low-dimensional node embeddings that faithfully reconstruct the original graph as well as explain a high percentage of variation in the expression data, suggesting that graph representation learning may aid the prediction of unobserved expression data.

\begin{table*}[]
\centering
\caption{Area under the ROC curve (AUC) and average precision (AP) for reconstructing biological network in \ecoli\ and mouse from five graph structure embedding approaches (detailed Section \ref{sec:graph-embedd-meth})}
\resizebox{1\columnwidth}{!}{%
\begin{tabular}{ccccccccc}
\hline
\multirow{3}{*}{Input}    & \multicolumn{4}{c}{AUC}                                                  & \multicolumn{4}{c}{AP}                                                 \\
                          & \multicolumn{3}{c}{Ecoli}                            & Mus Musculus      & \multicolumn{3}{c}{Ecoli}                            & Mus Musculus    \\
                          & TF\_net         & PPI              & Genetics        & PPI               & TF\_net         & PPI             & Genetics         & PPI             \\ \hline
Expression + Graph        & 0.868$\pm$0.017 & 0.8502$\pm$0.002 & 0.894$\pm$0.01  & 0.803$\pm$0.000   & 0.918$\pm$0.007 & 0.872$\pm$0.003 & 0.904$\pm$0.006  & 0.762$\pm$0.000 \\
Graph                     & 0.574$\pm$0.020 & 0.8316$\pm$0.018 & 0.882$\pm$0.014 & 0.6674$\pm$0.0000 & 0.727$\pm$0.012 & 0.860$\pm$0.010 & 0.904$\pm$0.006  & 0.653$\pm$0.000 \\
Expression + Random Graph & 0.529$\pm$0.01  & 0.4958$\pm$0.002 & 0.485$\pm$0.005 & 0.5108$\pm$0.0000 & 0.525$\pm$0.011 & 0.488$\pm$0.001 & 0.489$\pm$0.004  & 0.506$\pm$0.000 \\
Random Graph              & 0.492$\pm$0.006 & 0.4892$\pm$0.006 & 0.490$\pm$0.007 & 0.5025$\pm$0.0000 & 0.496$\pm$0.005 & 0.486$\pm$0.005 & 0.5002$\pm$0.003 & 0.505$\pm$0.000 \\
Expression                & 0.579$\pm$0.02           & 0.580$\pm$0.018  & 0.557$\pm$0.02  & 0.610$\pm$0.000   & 0.611$\pm$0.002 & 0.624$\pm$0.018 & 0.590$\pm$0.02   & 0.627$\pm$0.002
\end{tabular}
}
\label{tab:AUC}
\end{table*}

\begin{table*}[t]
\centering
\caption{The average Variance Explained on Gene Expression from the Regulatory Networks graph embedding}
\begin{tabular}{cccc}
\hline
\multirow{3}{*}{Input}    & \multicolumn{3}{c}{}               \\
                          & \multicolumn{3}{c}{Ecoli}                            \\
                          & TF\_net         & PPI             & Genetics         \\ \hline
Expression + Graph        & 0.883$\pm$0.005 & 0.805$\pm$0.003 & 0.7824$\pm$0.006 \\
Graph                     & 0.716$\pm$0.010 & 0.752$\pm$0.008 & 0.6697$\pm$0.010 \\
Expression + Random Graph & 0.660$\pm$0.010 & 0.667$\pm$0.013 & 0.6618$\pm$0.014 \\
Random Graph              & 0.647$\pm$0.017 & 0.652$\pm$0.018 & 0.6483$\pm$0.010
\end{tabular}
\label{tab:VE}
\end{table*}

\subsection{Predicting unobserved features for a fixed set of nodes}
\label{sec:pred-unobs-feat}

To investigate the use of graph neural networks in predicting unobserved node features, we first considered the problem where feature data is available for a constant set of training or predictor nodes in all experiments, in addition a set of training experiments is available with feature data for all graph nodes, and the task is to predict the values of the test nodes from test data for the training nodes (see Fig.~\ref{fig:train-test}(a)). As in Section~\ref{sec:approach}, we call the training data (for the train and test nodes) $\mathbf{X}$ the feature matrix and the testing data \textbf{Y} the label matrix, and hence the task is to predict the values of the test nodes in the label matrix.

We considered three categories of prediction methods: \textit{(i)} standard baseline methods that don't use graph information (linear regression, random forest regression, and a multi-layer perceptron, see Section~\ref{sec:pred-feat-from}), \emph{(ii)} standard regression methods trained on graph embeddings instead of directly on the training data (linear regression (LR) and random forest (RF) regression), and \emph{(iii)} graph feature auto-encoder methods for end-to-end learning of features (GCN, GraphSage, GraphConv, and FeatGraphConv, see Section~\ref{sec:mess-pass-neur}), see also Fig.~\ref{fig:auto-encoders}.

For the baseline models, we trained only on the expression values in the training data $\mathbf{X}$ and predicted the values of the test nodes for each dimension of $\mathbf{Y}$ separately.  For the indirect learning frameworks, first the graph embedding is obtained as in Section~\ref{sec:meas-relat-betw} using the graph auto-encoder, with and without using the training data $\mathbf{X}$ as node features, and then linear regression and random forest prediction models trained on the embedding in order to predict the values of the test nodes in \textbf{Y}, as in the baseline models. For each of the end-to-end graph convolution models, we trained the GNNs using the specified network and the training data $\mathbf{X}$ as node features and again predicted the values of the test nodes in \textbf{Y} in each dimension. Table \ref{tab:prediction} shows the performance (average mean squared error) of all methods on the \ecoli\ data. For this experiment the mouse single-cell RNA-seq data was omitted due to sparsity of the expression values.

As shown by the results, the newly proposed graph feature auto-encoder FeatGraphConv is able to predict the unobserved expression values better than the other graph convolution models, due to the fact that this layer is tailored to the prediction of features rather than the reconstruction of the graph. As expected, all end-to-end methods perform considerably better when training data is included as node features. The end-to-end methods, with the exception of GCN, also perform better than the indirect methods where regression models are trained on graph embeddings.

On the other hand, when the graph structure alone is used, the indirect embedding-based methods get less error. This could be due to the fact that these models better capture the structure of the graph, since their loss function is defined on the reconstruction of the adjacency matrix. Hence when the graph structure is the only information provided to the model, they are able to better capture this information and therefore obtain an embedding that better predicts expression data (on the basis of the results in Section~\ref{sec:meas-relat-betw}), compared to end-to-end models which try to predict the features directly and are operating blindly when no training features are provided.

We also observe that the lowest MSE overall is in fact obtained by baseline linear regression on the training data alone. However, since LR performs also better than MLP and RF in this context, we speculate that this result is due to the relative simplicity of the expression data with highly linear correlations among genes, and would not necessarily generalize to other data. Moreover, experiments on FeatGraphConv with 20,000 iterations (as opposed to the default of 500 used for all end-to-end methods in Table~\ref{tab:prediction}) showed that this model can decrease the MSE to $0.204\pm0.12$, $0.133\pm0.089$, and $0.107\pm0.083$ for each of the TF\_net, PPI, and Genetic networks, respectively. However, due to the high number of experiments and the need to train a different model for each of the experiments of \textbf{Y}, it is not computationally efficient to train the more complex GNN models with a higher number of iterations by default for this prediction task.

\begin{table*}[]
\centering
\caption{The average MSE of predicting test features of the test nodes using different models.}
\resizebox{1\columnwidth}{!}{%
\begin{tabular}{ccccccc}
\hline
\multirow{2}{*}{Method} & \multicolumn{2}{c}{TF\_Net}                          & \multicolumn{2}{c}{PPI}                             & \multicolumn{2}{c}{Genetics}                        \\
                        & Features                  & Graph                    & Features                 & Graph                    & Features                 & Graph                    \\ \hline
GCN                     & 7.791$\pm$3.550           & 15.127$\pm$2.280         & 6.208$\pm$0.607          & 11.106$\pm$0.52198       & 5.988$\pm$0.696          & 4.560$\pm$0.351          \\
GraphSAGE               & 0.332$\pm$0.160           & 8.078$\pm$2.592          & 0.265$\pm$0.135          & 2.844$\pm$0.349          & 0.233$\pm$0.127          & 4.466$\pm$1.605          \\
GraphConv               & 0.318$\pm$0.154           & 13.812$\pm$4.534         & 0.308$\pm$0.139          & 3.094$\pm$0.431          & 0.234$\pm$0.116          & 5.226$\pm$1.054          \\
FeatGraphConv           & \textbf{0.285$\pm$0.135}  & 7.525$\pm$2.941          & \textbf{0.244$\pm$0.130} & 5.207$\pm$1.476          & \textbf{0.201$\pm$0.112} & 3.414$\pm$0.691          \\ \hline
Multilayer perceptron   & 0.424$\pm$0.170           & -         & 0.354$\pm$0.153          & -         & 0.332$\pm$0.134          & -         \\
Linear Regression       & \textbf{0.21526+-0.12626} & -         & \textbf{0.147$\pm$0.105} & -          & \textbf{0.108$\pm$0.084} & -          \\
Random Forest           & 0.50969+-0.14335          & -         & 0.194 $\pm$0.103         & -          & 1.882$\pm$0.343          & -          \\ \hline
LR-embedding            & 1.583$\pm$0.200           & 2.279$\pm$0.403          & 1.091$\pm$0.166          & 1.453$\pm$0.264          & 1.863$\pm$0.332          & \textbf{1.653$\pm$0.271} \\
RF-embedding            & 1.945$\pm$0.31759         & \textbf{2.150$\pm$0.363} & 1.472$\pm$0.267          & \textbf{1.452$\pm$0.267} & 1.883$\pm$0.343          & 1.897$\pm$0.351         
\end{tabular}
}
\label{tab:prediction}
\end{table*}

\subsection{Predicting randomly distributed missing node features}

Based on the results in the previous section, we next considered the more challenging prediction task where unobserved node features are randomly distributed over the nodes and differ between experiments, that is, the task of imputing missing data (Fig.~\ref{fig:train-test}(b)). Since there are no fixed sets of training and test nodes, neither the baseline regression methods nor the indirect GNN learning frameworks are applicable in this case. In contrast, the end-to-end graph convolution methods allow to train a single model in an auto-encoder manner for the prediction of all the \textbf{Y} values, which may be placed in any possible order inside the feature matrix. We used these GNN models for the prediction of expression values in \ecoli\ and of non-zero values of the single-cell RNAseq data in mouse, and benchmarked them against two methods that don't use graph information, namely a normal MLP used in an auto-encoder scheme and MAGIC, a method designed specifically to impute missing data in single-cell RNA-seq data \cite{Magic} (see Section~\ref{sec:pred-feat-from}).

As shown in Table \ref{tab:imputation}, our FeatGraphConv convolution layer is able to predict missing features more accuratly compared to all other methods. It is interesting to note that the GNN models, with the exception of GCN, outperform MAGIC on the single-cell RNAseq imputation task, although the MLP, which does not use graph information, also performs well in this case.

\begin{table*}[]
\centering
\caption{The average MSE of predicting randomly distributed test values using different auto-encoder models.}

\begin{tabular}{cclclclc}
\hline
                       & \multicolumn{6}{c}{Ecoli}                                                                                                                   & Mus Musclus               \\ \hline
\multirow{2}{*}{Model} & \multicolumn{2}{c}{\multirow{2}{*}{TF\_Net}} & \multicolumn{2}{c}{\multirow{2}{*}{PPI}}     & \multicolumn{2}{c}{\multirow{2}{*}{Genetics}} & \multirow{2}{*}{PPI}      \\
                       & \multicolumn{2}{c}{}            & \multicolumn{2}{c}{}            & \multicolumn{2}{c}{}            &                           \\ \hline
GCN                    & \multicolumn{2}{c}{0.441$\pm$0.045}          & \multicolumn{2}{c}{0.314$\pm$0.019}          & \multicolumn{2}{c}{0.515$\pm$0.038}           & 0.083$\pm$0.313           \\
GraphSAGE              & \multicolumn{2}{c}{0.071$\pm$0.001}          & \multicolumn{2}{c}{0.056$\pm$0.0003}         & \multicolumn{2}{c}{0.059$\pm$0.001}           & 0.015$\pm$0.003           \\
GraphConv              & \multicolumn{2}{c}{0.144$\pm$0.003}          & \multicolumn{2}{c}{0.425$\pm$0.187}          & \multicolumn{2}{c}{0.487+-0.245}           & 0.015$\pm$0.001           \\
\textbf{FeatGraphConv} & \multicolumn{2}{c}{\textbf{0.059$\pm$0.002}} & \multicolumn{2}{c}{\textbf{0.049$\pm$0.002}} & \multicolumn{2}{c}{\textbf{0.044$\pm$0.001}}  & \textbf{0.010$\pm$0.0003} \\ \hline
MLP Auto-encoder       & \multicolumn{2}{c}{0.062$\pm$0.004}          & \multicolumn{2}{c}{0.053$\pm$0.004}          & \multicolumn{2}{c}{0.049$\pm$0.006}           & 0.013$\pm$0.002           \\
MAGIC                  & \multicolumn{2}{c}{3.505$\pm$0.006}          & \multicolumn{2}{c}{3.661$\pm$0.017}          & \multicolumn{2}{c}{3.215$\pm$0.003}           & 0.050$\pm$0.0002         
\end{tabular}
\label{tab:imputation}
\end{table*}

\section{Discussion}

In this paper we studied whether graph neural networks, which learn embeddings of nodes of a graph in a low-dimensional space, can be used to integrate discrete structures such as biological interaction networks with information on the activity of genes or proteins in certain experimental conditions. Traditionally, this is achieved by for instance network propagation methods, but these methods do not extract quantitative information from a graph that could be used for downstream modelling or prediction tasks. Graph neural networks on the other hand can include node features (gene or protein expression levels) in the learning process, and thus in theory can learn a representation that better respects the information contained in both data types. Thus far the integration of node features in graph representation learning has mainly been pursued for the task of link prediction. Here instead we focused on the task of predicting unobserved or missing node feature values.

We showed that representations learned from a graph and a set of expression profiles simultaneously result in better reconstruction of the original graph and higher expression variance explained than using either data type alone, even when the representations are trained on the graph reconstruction task. We further proposed a new end-to-end graph feature auto-encoder which is trained on the feature reconstruction task, and showed that it performs better at predicting unobserved node features than auto-encoders that are trained on the graph reconstruction task before learning to predict node features.

Predicting or imputing unobserved node features is a common task in bioinformatics. In this paper we demonstrated the value of our graph feature auto-encoder on the problem of imputing missing data in single-cell RNAseq data, where it performs better than a state-of-the-art method that does not include protein interaction data. Other potential application areas are the prediction of new disease-associated genes from a seed list of known disease genes on the basis of network proximity \cite{cowen2017network}, or the prediction of non-measured transcripts or proteins from new low-cost, high-throughput transcriptomics and proteomics technologies that only measure a select panel of genes or proteins \cite{subramanian2017next,suhre2020genetics}.

A potential drawback of our method is that it assumes that the interaction graph is known and of high-quality. Future work could investigate whether it is feasible to learn graph representations that can do link prediction and node feature prediction simultaneously, or whether network inference followed by graph representations learning for one type of omics data (e.g. bulk RNAseq data) can aid the prediction of another type of omics data (e.g. single-cell RNAseq).

In summary, graph representation learning using graph neural networks is a powerful approach for integrating and exploiting the close relation between molecular interaction networks and functional genomics data, not only for network link prediction, but also for the prediction of unobserved functional data.

\cleardoublepage{}
\renewcommand\thesection{A\arabic{section}}
\renewcommand\thesubsection{A.\arabic{subsection}}
\renewcommand\thefigure{A\arabic{figure}}
\renewcommand\thetable{A\arabic{table}}
\renewcommand\theequation{A\arabic{equation}}
\setcounter{figure}{0}
 \setcounter{table}{0}
 \setcounter{section}{0}
 \setcounter{equation}{0}

\section*{Appendix}
\subsection{Variance explained on features}
\label{sec:vari-expl-feat}
For an embedding matrix $\Z\in\R^{N\times H}$ and feature matrix $\X\in\R^{N\times M}$, we calculated the amount of variance in $\X$ explained by $\Z$ as
\begin{equation}
  \label{eq:var-expl}
  \mathbb{V}_\Z = \frac{\tr\bigl(\P_\Z\X^T\X\bigr)}{\tr(\X^T\X)}
\end{equation}
where $\P_\Z$ is the projection matrix onto the subspace of $\R^N$ spanned by the columns of $\Z$,
\begin{align*}
  \P_\Z = \Z (\Z^T\Z)^{-1}\Z^T.
\end{align*}
Note that if we write the eigendecomposition of $\X^T\X$ as $\X^T\X=\V^T\Dbf\V$, then the columns of $\V$ corresponding to the nonzero eigenvalues in $\Dbf$ are the principal components of $\X$. If $\Z$ consist of the $i^{th}$ principal component, then eq.~(\ref{eq:var-expl}) reduces to the familiar variance explained by this component, $\Delta_i/(\sum_j\Delta_j)$. If $\Z$ consists of a single vector $z\in\R^N$ with unit length, $\|z\|=1$, then eq.~(\ref{eq:var-expl}) reduces to
\begin{align*}
  \mathbb{V}_z = \sum_{i=1}^N \frac{\Delta_i}{\sum_j\Delta_j} (z^Tu_i)^2,
\end{align*}
a weighted sum of the variances explained by each principal component, with weights determined by the extent of overlap between $z$ and each principal component. Eq.~(\ref{eq:var-expl}) generalizes this to summing the variances explained by multiple vectors simultaneously that need not be mutually orthogonal.
\subsection{Random Graph Generation using Erd\H{o} s–R\'enyi model}
\label{sec:rand-graph-gener}
According to Erd\H{o} s–R\'enyi, a random graph $G(n,p)$ has ${n\choose 2}p $ edges placed at random.  The degree distribution of each node is calculated through:
\begin{equation}
    P(\degr(v)=k)={n-1\choose k}p^k(1-p)^{n-1-k},
\end{equation}
which is the binomial distribution. We adjusted the value of $p$ so that the random graphs would approximately have the same number of edges as the real networks.
\end{document}